\documentstyle[aps,amssymb]{revtex}

\begin{document}
\title{Role of degenerate atomic levels in the entanglement and the decoherence }
\author{L. Zhou\thanks{%
E-mail:zhlhxn@dlut.edu.cn}, H. S. Song\thanks{%
E-mail:hssong@dlut.edu.cn}, Y. X. Luo}
\address{{\small Department of Physics, Dalian University of Technology, Dalian,}\\
116023, P. R. China.}
\maketitle

\begin{abstract}
We studied the dispersive dissipation of denegerate-level atom interacting
with a single linearly-polarized mode field. It is found that the degeneracy
of the atomic level affects the dissipation bahavior of the system as well
as the subsystems. The degeneracy of the atomic level augment the periods of
entanglement and increase the degree of the maxima statistical mixture
states.

\end{abstract}

\baselineskip=18.2pt

\section{Introduction}

Since the EPR paradox was proposed, the quantum entanglement has been a
interesting subject, which reveal the profound difference between quantum
and classical world. Recently, entanglement as a physical resource has been
used in quantum information such as quantum teleportation, superdense coding
and quantum cryptography [1,2,3]. Another equally fundamental question
concerns the nonexistence of coherence superposition of macroscopically
distinguishable states, illustrated by the Schr\"{o}dinger's cat paradox.
One of the answer to the second question stress the role of dissipation on
the disappearance of coherence [4,5]. Decoherence follows from the
irreversible coupling of the observed system to the outside word reservoir,
and this coupling induce that decoherence of the macroscopic states would be
too fast to be observed.

A atom (atoms) interaction with quantum electromagnetic field play important
role in explaining these essential quantum problem, in preparation some
kinds of quantum states [6,7] and in monitoring decoherence [8]. Several
schemes had been proposed to generate the entangle atomic state, on the
condition that the entanglement between the first atom and the cavity field
can survive for long enough so that it can be transferred to a second atom
via coherent interaction [6,7]. The superposition of two coherent states can
be prepared and decoherence can be monitored in cavity QED. These schemes
concerned with a atom or atoms interacting with field in a cavity, and the
dissipation of the cavity play important role in the entanglement of
subsystem and in the decoherence of the system or subsystem [9,10,17].
Therefore, in Ref. [11] the dispersive atomic evolution in a
dissipative-driven cavity was studied, and the influence of dissipation on
the entanglement and on the decoherence was investigated via JCM in the
dispersive approximation in Ref. [12].

Although theoretical predictions based on the simple two-level model have
proven to be powerful, pure two-level systems are seldom found in real
experiments. In most cases, the atomic level are degenerate [13,14]. If the
levels of an isolated atom are degenerate in the projection of the total
electronic angular momenta on the quantization axis, we should take into
account the degeneracy of atomic level [13]. In Ref.[15,16] original
Jaynes-Cummings model (JCM) was generalized to the case of degenerate atomic
levels. If results of degeneracy of atomic level could provide some
available properties, we may turn the pure two-level atom into degenerate
atom by introducing magnetic field. In this paper, we study dynamics of a
degenerate atom interacting with the field in a dissipative cavity. In
dispersive approximation, we find that the degeneracy of atomic level
augment the period of disentanglement between the atom and the field and
increase the degree of the maxima statistical mixture states.

\section{The Jaynes-Cummings model with degenerate atomic level and the
dispersive approximation}

Let us take into account the degeneracy of atomic levels, the full set of
states of the system may be written as

\begin{equation}
|n,J_{_\alpha },m_{_\alpha }>=|n>\cdot |J_{_\alpha },m_{_\alpha
}>,n=0,1,...,m_{_\alpha }=-J_{_\alpha },...J_{_\alpha ,}\alpha =b,c,
\end{equation}
where $n$ is the number of photons in the field mode, while $b$ and $c$
denote the upper and lower atomic levels respectively. $J_b$ and $J_c$ are
the values of the total electronic angular momenta of resonant levels, while 
$m_b$ and $m_c$ are their projections on the quantization axis, the
Cartesian axis Z, which is directed along the polarization vector of the
field mode.

We assume that a degenerate-level atom interact with a single
linearly-polarized mode field. Hence only that atomic transition could emit
a linearly-polarized photon take part in the interaction. The Hamiltonian of
the system may be written as ($\hslash =1$) 
\begin{equation}
H=\omega _0a^{+}a+\frac 12\omega (n_b-n_c)+g(a^{+}S_{-}+aS_{+}),
\end{equation}
where $\hat{a}^{+}$and $\hat{a}$ are the operators of the creation and
annihilation of photons with frequency $\omega _{0\text{ }}$in the field
mode, and 
\begin{equation}
n_{_\alpha }=\sum_{m_\alpha =-J_\alpha }^{J_\alpha }|J_{_\alpha },m_{_\alpha
}\rangle \langle J_{_\alpha },m_{_\alpha }|,\qquad \alpha =b,c,
\end{equation}
are the operators of total population of resonant atomic levels $b$ and $c$, 
$\omega $ is the frequency of the optically-allowed atomic transition $%
J_b\longrightarrow J_c.$%
\begin{equation}
S_{-}=\sum_m\alpha _m|J_{c,}m\rangle \langle J_b{}_{,}m|
\end{equation}
is the dipole moment operator of the atomic transition $J_b\longrightarrow
J_c$, where 
\begin{equation}
\alpha _m=(-1)^{j_b-m}\left( 
\begin{array}{lll}
J_b & 1 & J_c \\ 
-m & 0 & m
\end{array}
\right)
\end{equation}
is matrix elements defined through Wigner 3j-symbol, corresponding to the
linearly-polarized photon.

We consider the far-off resonance limit for the atom-field interaction
(dispersive interaction ). The Hamiltonian take the form 
\begin{equation}
H_I=\frac \delta 2(n_b-n_c)+g(a^{+}S_{-}+aS_{+}),
\end{equation}
where the detuning $\delta =\omega -\omega _0$ . The dipole moment operator
of the atomic transition $S_{-}$ and $S_{+\text{ }}$satisfy the commutation
relation

\qquad 
\begin{equation}
\lbrack S_{+},S_{-}]=2S_z,\qquad [S_{\pm },S_z]=\pm S_z,
\end{equation}
with 
\begin{equation}
S_z=\frac 12\sum_m\alpha _m^2(|J_{b,}m\rangle \langle
J_b{}_{,}m|-|J_{c,}m\rangle \langle J_c{}_{,}m|).
\end{equation}

To solve the master equation in next section, here we take a set of unitary
transformation to the Hamiltonian of Eq.(6) which is proposed in Ref.[11].
This transformation correspond to small rotation in the SU(2) group with an
operator parameter 
\begin{equation}
H_{eff}=U_2U_1HU_1^{+}U_2^{+},
\end{equation}
where 
\[
U_1=\exp (i\frac{\sqrt{2}g}\delta pS_x),\qquad U_2=\exp (i\frac{\sqrt{2}g}%
\delta qS_y), 
\]
and 
\[
q=\frac 1{\sqrt{2}}(a+a^{+}),\qquad p=\frac i{\sqrt{2}}(a^{+}-a), 
\]

\[
S_x=\frac 12(S_{+}+S_{-}),\qquad S_y=-\frac i2(S_{+}-S_{-}). 
\]
Keeping terms up to first order in $\sqrt{2}g/\delta \ll 1$, we get 
\begin{equation}
H_{eff}=\frac \delta 2(n_b-n_c)+\frac{g^2}\delta (R_b+R_c)+\frac{g^2}\delta
(2a^{+}a+1)S_z,
\end{equation}
where 
\begin{equation}
R_{_\alpha }=\sum_m\alpha _m^2(|J_{\alpha ,}m\rangle \langle J_{_\alpha
}{}_{,}m|,\qquad \alpha =b,c.
\end{equation}

Note that the fore two terms of Eq. (10) commute with the effective
Hamiltonian $H_{eff}$. We can further simplify the effective Hamiltonian by
the following transformation of the operator $f$:

\begin{equation}
\tilde{f}=e^{i[\frac \delta 2(n_b-n_c)+\frac{g^2}\delta (R_b+R_c)]}fe^{-i[%
\frac \delta 2(n_b-n_c)+\frac{g^2}\delta (R_b+R_c)]}.
\end{equation}
Thus, we finally get 
\begin{equation}
\tilde{H}_{eff}=\Omega (2a^{+}a+1)S_z,
\end{equation}
with $\Omega =\frac{g^2}\delta $. In the next section, we will directly
using the expression of Eq.(13).

\section{The master equation and its solution}

We assume that there is a reservoir coupled to the field in the usual way.
Using the transformation of Eq. (12) to the density matrix, master equation
has a standard form 
\begin{equation}
\frac{d\tilde{\rho}}{dt}=i[\tilde{\rho},\tilde{H}_{eff}]+{\cal D}\tilde{\rho}%
.
\end{equation}
The losses in the cavity are phenomenologically represented by the
superoperator ${\cal D}$. At the zero temperature, we have 
\begin{equation}
{\cal D}\tilde{\rho}=\kappa (2a\tilde{\rho}a^{+}-a^{+}a\tilde{\rho}-\tilde{%
\rho}a^{+}a),
\end{equation}
where $\kappa $ is the damping constant. The density operator $\tilde{\rho}$
belongs to the set $\Re $ ($\Re _A\otimes \Re _F$) of the trace class
operators that act in the space corresponding to the direct product of the
two Hilbert space $\Re _A$ and $\Re _F$ of the atom and the field,
respectively. We can represent the density operator as following : 
\begin{equation}
\tilde{\rho}=\sum_{m,m^{\prime },J_{_\alpha },J_{_{_\beta }}}\tilde{\rho}%
_{_{J_{_\alpha }J_{_{_\beta }}mm^{\prime }}}|J_{_\alpha },m\rangle \langle
J_{_{_\beta }},m^{\prime }|,\text{ with }J_{_\alpha }(J_{_{_\beta }})=J_b,J_c
\end{equation}
where 
\[
\tilde{\rho}_{_{J_{_\alpha }J_{_{_\beta }}mm^{\prime }}}=\langle J_{_\alpha
},m|\tilde{\rho}|J_{_{_\beta }},m^{\prime }\rangle 
\]

The formal solution of the master equation (14) is given by 
\begin{equation}
\tilde{\rho}(t)=e^{{\cal L}t}\rho (0).
\end{equation}
we employ superoperators language which is employed in Refs.[12,20,21].
Substituting Eq.(16) into Eq.(14), we obtained Liouvillians corresponding to
the matrix elements $\tilde{\rho}_{_{J_{_b}J_bmm^{\prime }}}$ and $\tilde{%
\rho}_{_{J_{_c}J_cmm^{\prime }}}$ as

\begin{equation}
{\cal L}_{J_{_\alpha }J_{_{_\alpha }}mm^{\prime }}=\mp 2i(\omega _m{\cal M}%
-\omega _{m^{\prime }}{\cal P})+\kappa (2{\cal F}-{\cal M}-{\cal P})\mp
i(\omega _m-\omega _{m^{\prime }}),
\end{equation}
where $\omega _m=\alpha _m\Omega $. We choose $-$ when we calculate $\tilde{%
\rho}_{_{J_{_b}J_bmm^{\prime }}}$ and $+$ corresponding to $\tilde{\rho}%
_{_{J_{_c}J_cmm^{\prime }}}$. The superoperators in Eq.(18) are defined as $%
{\cal \digamma }\hat{\rho}=\hat{a}\hat{\rho}\hat{a}^{+},{\cal M}\hat{\rho}=%
\hat{a}^{+}\hat{a}\hat{\rho},{\cal P}\hat{\rho}=\hat{\rho}\hat{a}^{+}\hat{a}$%
. They satisfy the commutation relation 
\begin{equation}
\lbrack {\cal \digamma },{\cal M}]={\cal \digamma },\qquad [{\cal \digamma },%
{\cal P}]={\cal \digamma },\qquad [{\cal M},{\cal P}]=0.
\end{equation}
In the same way, we also get the Liouvillian of acting in $\tilde{\rho}%
_{_{J_{_b}J_cmm^{\prime }}}$ as 
\begin{equation}
{\cal L}_{J_bJ_cmm^{\prime }}=-2i(\omega _m{\cal M}+\omega _{m^{\prime }}%
{\cal P})+\kappa (2{\cal F}-{\cal M}-{\cal P})-i(\omega _m+\omega
_{m^{\prime }}).
\end{equation}
A similar expression of Liouvillian ${\cal L}_{JcJ_bmm^{\prime }}$ which act
in $\tilde{\rho}_{_{J_cJ_bmm^{\prime }}}$ is easy obtained except for taking
conjugate of Eq.(20).

\section{Time evolution of initial state}

We assume the initial state of the system as 
\begin{equation}
\Psi (0)=\frac 1{\sqrt{2}}\sum_m(%
{\textstyle {1 \over \sqrt{2J_b+1}}}%
|J_b,m\rangle \langle J_b,m|+%
{\textstyle {1 \over \sqrt{2J_c+1}}}%
|J_c,m\rangle \langle J_c,m|)\otimes |\alpha \rangle
\end{equation}
the atom is in the degenerate level in equal probability and it enters the
cavity in a coherence superposition and finds there a coherent field state $%
|\alpha \rangle $ , therefore initially

\[
\tilde{\rho}_{_{J_{_b}J_bmm^{\prime }}}(0)=%
{\textstyle {1 \over 2(2J_b+1)}}%
|\alpha \rangle \langle \alpha |,\tilde{\rho}_{_{J_{_c}J_cmm^{\prime }}}(0)=%
{\textstyle {1 \over 2(2J_c+1)}}%
|\alpha \rangle \langle \alpha |, 
\]

\[
\tilde{\rho}_{_{J_{_b}J_cmm^{\prime }}}(0)=\tilde{\rho}_{_{J_{_c}J_bmm^{%
\prime }}}(0)=%
{\textstyle {1 \over 2\sqrt{(2J_b+1)(2J_c+1)}}}%
|\alpha \rangle \langle \alpha |. 
\]
Solving the Eq. (17), we finally get the density matrix 
\begin{eqnarray}
\tilde{\rho} &=&\frac 12\sum_{m,m^{\prime }}\{%
{\textstyle {1 \over 2J_b+1}}%
\exp [\Gamma (\chi _{mm^{\prime }},t)+i\Theta (\chi _{mm^{\prime
}},t)]|J_b,m,\alpha (t)e^{-2i\omega _mt}\rangle \langle J_b,m^{\prime
},\alpha (t)e^{-2i\omega _{m^{\prime }}t}|  \nonumber \\
&&+%
{\textstyle {1 \over 2J_c+1}}%
\exp [\Gamma (\chi _{mm^{\prime }},t)-i\Theta (\chi _{mm^{\prime
}},t)]|J_c,m,\alpha (t)e^{2i\omega _mt}\rangle \langle J_c,m^{\prime
},\alpha (t)e^{2i\omega _{m^{\prime }}t}|  \nonumber \\
&&+%
{\textstyle {1 \over \sqrt{(2J_b+1)(2J_c+1)}}}%
[\exp (\Gamma (\lambda _{mm^{\prime }},t)+i\Theta _m(\lambda _{mm^{\prime
}},t))|J_b,m,\alpha (t)e^{-2i\omega _mt}\rangle \langle J_c,m^{\prime
},\alpha (t)e^{2i\omega _{m^{\prime }}t}|  \nonumber \\
&&+\exp (\Gamma (\lambda _{mm^{\prime }},t)-i\Theta _m(\lambda _{mm^{\prime
}},t))|J_c,m,\alpha (t)e^{2i\omega _mt}\rangle \langle J_b,m^{\prime
},\alpha (t)e^{-2i\omega _{m^{\prime }}t}|]\}  \label{p}
\end{eqnarray}
where $\chi _{mm^{\prime }}=\omega _m-\omega _{m^{\prime }},\lambda
_{mm^{\prime }}=\omega _m+\omega _{m^{\prime }}.$ 
\begin{equation}
\Gamma (x,t)=-|\alpha |^2(1-e^{-2\kappa t})-\frac{|\alpha |^2\kappa }{\kappa
^2+x^2}[e^{-2\kappa t}(\kappa \cos 2xt-x\sin 2xt)-\kappa ],
\end{equation}
and 
\begin{equation}
\Theta (x,t)=-xt+\frac{|\alpha |^2\kappa }{\kappa ^2+x^2}[e^{-2\kappa
t}(x\cos 2xt+\kappa \sin 2xt)-x],
\end{equation}
where $x$ equal to $\chi _{mm^{\prime }}$ and $\lambda _{mm^{\prime }}$,
respectively. The function $\Gamma (x,t)$ in Eq. (23) embody the effect of
reservoir because it vanishes for $k\rightarrow 0$ .

The coherence properties of this density operator as a function of time is
conveniently studied by means of the linear entropy 
\begin{equation}
S=1-Tr(\rho ^2).  \label{s}
\end{equation}
The quantity $Tr(\rho ^2)$ can be taken as a measure of the degree of purity
of the reduced state; for a pure state $S$ is zero but for $0\prec S\preceq
1 $ the state corresponds to a mixture, with information effectively lost.
Because the all the transformation in Eq.(9) and in Eq.(12) are unitary,
hence the entropy is not affected by the transformation. Hereafter we will
direct use these density operators ( $\tilde{\rho}_F$, $\tilde{\rho}_A$) to
gain corresponding entropy. The linear entropy of the total system is
obtained from Eq.(22)

\begin{equation}
S=1-\frac 14\sum_{m,m^{\prime }}\{[%
{\textstyle {1 \over (2J_b+1)^2}}%
+%
{\textstyle {1 \over (2J_c+1)^2}}%
]\exp [2\Gamma (\chi _{mm^{\prime }},t)]+%
{\textstyle {2 \over (2J_b+1)(2J_c+1)}}%
\exp [2\Gamma (\lambda _{mm^{\prime }},t)]\}.
\end{equation}
Note that the coherence properties of the total system is also completely
governed by the presence of the reservoir, denoted by the function $\Gamma
(x,t)$. This is similar to the usual dissipation of JCM [12]. However the
linear entropy is the sum of $^{``}$m$^{"}$ which is related to the value of 
$J_b$ and $J_c$, angular momenta of the two atomic level. This difference
would result in some marvelous novel properties. In the succeeding section
we will numerate some results and compare these novel properties with that
of Ref. [12].

Taking now the trace of the global density $\tilde{\rho}$ on the atomic
variables, we get the reduced field density 
\begin{equation}
\tilde{\rho}_F=\frac 12\sum_m[%
{\textstyle {1 \over 2J_b+1}}%
|\alpha (t)e^{-2i\omega _mt}\rangle \langle \alpha (t)e^{-2i\omega _mt}|+%
{\textstyle {1 \over 2J_c+1}}%
|\alpha (t)e^{2i\omega _mt}\rangle \langle \alpha (t)e^{2i\omega _mt}|].
\end{equation}
The linear entropy of the field is obtained by 
\begin{eqnarray}
S_F &=&1-\frac 14\sum_{m,m^{\prime }}\{[%
{\textstyle {1 \over (2J_b+1)^2}}%
+%
{\textstyle {1 \over (2J_c+1)^2}}%
]\exp (-4|\alpha (t)|^2\sin ^2\chi _{mm^{\prime }}t)  \nonumber \\
&&+%
{\textstyle {2 \over (2J_b+1)(2J_c+1)}}%
\exp (-4|\alpha (t)|^2\sin ^2\lambda _{mm^{\prime }}t)\}.
\end{eqnarray}
Note also that although it is the field, which is directly coupled to the
reservoir, the function $\Gamma (x,t)$ , characteristic function of this
coupling, does not appear in the linear entropy of the field but of the
atom. In order to analyze what happens to the atom, we trace out the field
variables from Eq. (22) and get

\begin{eqnarray}
\tilde{\rho}_A &=&\frac 12\sum_{m,m^{\prime }}\{%
{\textstyle {1 \over 2J_b+1}}%
\exp [\Gamma (\chi _{mm^{\prime }},t)+i\Theta (\chi _{mm^{\prime
}},t)-|\alpha (t)|^2(1-e^{-2i\chi _{mm^{\prime }}t})]|J_b,m\rangle \langle
J_b,m^{\prime }|  \nonumber \\
&&+%
{\textstyle {1 \over 2J_c+1}}%
\exp [\Gamma (\chi _{mm^{\prime }},t)-i\Theta (\chi _{mm^{\prime
}},t)-|\alpha (t)|^2(1-e^{2i\chi _{mm^{\prime }}t})]|J_c,m\rangle \langle
J_c,m^{\prime }|  \nonumber \\
&&%
{\textstyle {1 \over \sqrt{(2J_b+1)(2J_c+1)}}}%
[\exp (\Gamma (\lambda _{mm^{\prime }},t)+i\Theta _m(\lambda _{mm^{\prime
}},t)-|\alpha (t)|^2(1-e^{-2i\lambda _{mm^{\prime }}t})|J_b,m\rangle \langle
J_c,m^{\prime }|  \nonumber \\
&&+\exp (\Gamma (\lambda _{mm^{\prime }},t)-i\Theta _m(\lambda _{mm^{\prime
}},t)-|\alpha (t)|^2(1-e^{2i\lambda _{mm^{\prime }}t}))|J_c,m\rangle \langle
J_b,m^{\prime }|]\}.
\end{eqnarray}
Atomic coherence loss will be measured by its linear entropy 
\begin{eqnarray}
S_A &=&1-\frac 14\sum_{m,m^{\prime }}\{[%
{\textstyle {1 \over (2J_b+1)^2}}%
+%
{\textstyle {1 \over (2J_c+1)^2}}%
]\exp (2\Gamma (\chi _{mm^{\prime }},t)-4|\alpha (t)|^2\sin ^2\chi
_{mm^{\prime }}t)  \nonumber \\
&&+%
{\textstyle {2 \over (2J_b+1)(2J_c+1)}}%
\exp (2\Gamma (\lambda _{mm^{\prime }},t)-4|\alpha (t)|^2\sin ^2\lambda
_{mm^{\prime }}t)\}.
\end{eqnarray}
The coherence of the atom is determined by the dissipative cavity (denoted
by the $\Gamma (x,t)$ function) as well as the entanglement (proportional to 
$|\alpha |^2$). Most important thing is that the degenerate atomic level
take effect.

\section{Results and discussion}

The levels b and c in the experiments [18,19] were Rydberg states of the
rubidium atom with the angular momenta $J_b=\frac 32$ and $J_c=\frac 32$ or $%
J_c=\frac 52$. Here we take $J_b$ and $J_c$ both are $\frac 32$ , in this
case, 
\begin{equation}
\alpha _{\frac 12}=\alpha _{-\frac 12}=\frac 1{2\sqrt{15}},
\end{equation}
and 
\begin{equation}
\alpha _{_{\frac 32}}=\alpha _{_{-\frac 32}}=\frac 3{2\sqrt{15}}.
\end{equation}
According to Eq. (28), we plot the evolution of the field's linear entropy.
Note that the behavior of the coherence loss of field is not sine
oscillation but we still observe that the field exhibit periodic
disentanglement. As disentanglement take place, the field is in a pure
state, corresponding to $S_F(t_d)=0$. However this period $t_d$ are much
longer than $t_d^{^{\prime }}=\frac \pi \Omega $ which are the case in the
usual dissipative JCM in dispersive approximation [12]. With the parameter
of our choice, the entanglement period $t_d=\frac{12.2}\Omega $ . In other
word, the entanglement can survive for long time. Comparing the maxima
values of $S_F$, corresponding to the maxima degree of mixture state, with
that of in Ref. [12], we surprisingly find that the maxima values of $S_F$
are greater than o.5, the characteristic values of two statistical mixture
states.

If one carefully examine the form of $\tilde{\rho}_{_F}$ in Eq. (27), one
can see that the field are mixture of all kinds of states $|\alpha (t)e^{\pm
i\omega _mt}\rangle $. Thus the maxima degree of mixture state relate to the
values of $"m"$, and maxima values of $S_F$ are larger than o.5. In usual
dissipative JCM, the field are mixture of the two state $|\alpha (t)e^{\pm
i\omega t}\rangle $, hence the maxima values of $S_F$ equal to 0.5. On the
other hand, different $\omega _m$ correspond to different periods, the
result of summation should take the minimum common multiple. So we can
observe the longer period of entanglement. Therefore, on one hand, the
degenerate atomic level increase the period of entanglement, on the other
hand, it enhance the degree of maxima mixture state.

To verify the role of degeneracy of atomic level and dissipation on the
coherence loss of atom and the system, we show $S(t)$ and $S_A(t)$ as a
function of time for two values of $\kappa $. We observe that the larger
dissipation, the more rapid of the coherence loss of the atom and the
system. When the atom and the field disentangle, the field is in a pure
state, the atom carries alone the degree of the decoherence of the system.
At the instants of disentanglement $S(t_d)=S_A(t_d)$, while $S_F(t_d)=0$.
This property is the same as in the general JCM without the dissipation. We
also find that the role of the degeneracy atomic level is to increase the
period of entanglement and disentanglement. This is coincide to the Fig.
(1). However, the increased periods of entanglement have nothing to do with
the dissipation. In Fig. (3), we draw the evolution of the linear entropy of
atom and the system alone with the intensity of the cavity. It is clear that
the periods of the entanglement are not relate to the intensity of cavity.
Note that the asymptotic value of $S(t)$ and $S_A(t)$ grow with the
intensity and the asymptotic value break through the asymptotic limits $%
\frac 12$, the characteristic of the statistical mixture. On the other hand,
with the increase of the intensity of the cavity, the atom and the system
lost their purity more rapidly.

\section{Conclusion}

Taking into account the degeneracy of atomic level, we studied the
dissipation of degenerate atom interaction with a single linearly-polarized
mode field in dispersive approximation. The degeneracy of the atomic level
affect the dissipation behavior of the system as well as the subsystems. We
find that the degeneracy of the atomic level augment the period of
entanglement between the atom and the field and increase the degree of the
maxima statistical mixture states.

It is worthwhile to point out that the available of the augmented period of
entanglement. The entanglement as a physical resource is available on the
condition that the entanglement could keep long enough so that we can
accomplish some task. For example, in Ref. [12] as we mention before, the
entanglement between the first atom and the cavity field must survive long
enough so as to generate the entanglement atomic state. At this point, the
large period of entanglement have some advantage, although the entanglement
state become complicated.

\section{Acknowledgment}

The work is supported by the Chinese Education Foundation through grant No.
1999014105.

\bigskip 

The caption of the figures:

Fig. 1 The evolution of the field's linear entropy where $|\alpha |^2=1.0$, $%
\kappa /\Omega =0.01$.

Fig.2 The linear entropy of the systems (solid line ) and of the atom (dot
line) as a function of $\Omega t$.

Fig. 3 Linear entropy of the systems (solid line ) and of the atom (dot
line) as a function of amplitude $\alpha $ where $\kappa /\Omega =0.02$.

\end{document}